\documentstyle[psbox]{EAYAM2006}
\newcommand{\HII}{H\,{\sc ii}}
\newcommand{\sun}{\odot}

\newcommand{\arcsec}{''}

\begin{document}

\title{High Resolution Radio Maps of Four Nearby Spiral Galaxies}

% AUTHOR(S) 
\author{
Chao-Wei Tsai$^1$, Jean L. Turner$^1$, Sara C. Beck$^2$, \\ 
Lucian P. Crosthwaite$^3$, Paul T. P. Ho$^{4}$, and David S. Meier$^{5}$
\\[12pt]  % TO BE SPACED WITH ONE LINE
%
% INSTITUTES OF AUTHORS
$^1$  Department of Physics and Astronomy, UCLA, Los Angeles, CA 90095-1547, U.S.A. \\
$^2$  Department of Physics and Astronomy, Tel Aviv University, Ramat Aviv, Israel \\
$^3$  Northrop Grumman, San Diego, CA, U.S.A. \\
%$^4$  Harvard-Smithsonian Center for Astrophysics, Cambridge, MA, U.S.A. \\
$^4$  Institute of Astronomy and Astrophysics, Academia Sinica, Taipei, Taiwan \\
$^5$  National Radio Astronomy Observatory, Socorro, NM, U.S.A.\\
%$^7$  Department of Astronomy, University of Illinois, Urbana, IL, U.S.A.\\
%
% please put the first author's initial and e-mail address below
{\it E-mail(C.-W.T.): cwtsai@astro.ucla.edu} 
%            \_ Initial      \
%                             \_ E-mail address
}

\abst{
We report subarcsecond-resolution VLA imaging of the centers of four nearby spiral galaxies: IC 342, Maffei II, NGC 2903, and NGC 6946. In each galaxy, 7 - 12 compact radio continuum sources were identified within the central $\sim 15'' \times 15''$. 
%These compact sources are responsible for 20 - 30$\%$ of the total emission from the central kpc of the host galaxies at 2 cm, but only 5 - 9$\%$ at 6 cm, except in NGC 6946, which has compact non-thermal sources. 
Slightly over half of the compact sources appear to be \HII\ regions with flat or positive spectral indices ($\alpha \gsim -0.1$). The \HII\ regions with rising spectra are optically thick at centimeter wavelengths, and thus dense ($n_{i} \sim 10^{4}~cm^{-3}$) and young. The largest of these \HII\ regions require the excitation of 500 - 800 O stars, within regions of only few parsecs extent. These clusters approach the sizes expected for globular clusters. %The thermal sources are preferentially projected to be near the centers of giant molecular clouds. Most of the thermal compact sources do not have obvious counterparts in the visual bands due to high extinction.
}

%\kword{workshop: proceedings --- LaTeX2.09: style file --- instructions}

\maketitle
\thispagestyle{empty}

\section{Introduction}

The formation of open clusters has been well studied in the past two decades (see review by Lada \& Lada 2003). In contrast, due to the lack of well-studied young clusters as massive as a typical globular cluster, little is known about the formation of globular clusters (Larson 1992). Although it has been suggested that globular clusters share some basic formation mechanism with open clusters due to the apparent continuity in properties (Larsen 1992), the formation of two types clusters with few orders of magnitude difference in total mass might be induced by significantly different physical phenomena. It has been suggested that the young clusters found by \textit{Hubble Space Telescope} (\textit{HST}) are protoglobular clusters formed by merger events (Holtzman et al. 1992; and the review by Whitmore 2003). However, how the star formation is triggered by mergers and what triggers the star cluster formation in non-interacting galaxies (B{\" o}ker et al. 2002) are still unclear.

In order to understand the formation of globular clusters, it is essential to trace the youngest protoglobular clusters. These young and massive star clusters, or super star clusters -- ``SSC'', contain hundreds to thousands O stars in their very first few million years of life. They are often embedded in their dusty natal cocoons  (Wynn-Williams et al. 1972) and visually obscured by their surrounding birth clouds. The high optical extinction is true as well for the \HII\ regions which are large enough, with much of the extinction internal to the nebula itself (Kawara et al. 1989; Ho et al. 1990; Beck et al. 1996). Thus, the forming SSC usually cannot be seen in optical. However, embedded clusters can be detected through their nebulae, which glow by reprocessed UV light from O stars. The thermal free-free emission from these nebulae can be detected and identified at radio wavelengths. Hence radio continuum observations which suffer much less extinction are often useful tracers of young star forming regions.

We report the centimeter continuum study at 2 and 6 cm on the centers of four nearby and well-studied spiral galaxies, IC 342, Maffei II, NGC 2903, and NGC 6946. The centers of these galaxies are infrared-bright and molecular gas rich. The goal of this investigation is to identify young SSC candidates in the star formation regions in the centers of these galaxies using subarcsecond radio continuum imaging. Subarcsecond imaging with the \textit{VLA} in its extended configuration maximizes sensitivity to bright and compact radio ``supernebulae'' over low brightness disk synchrotron emission. Observing at shorter wavelengths, $\lambda<$ 2 cm, also minimizes the contribution of synchrotron emission, which falls with frequency.

The results presented here are detailed in Tsai et al. (2006)

\begin{table*}
\caption{\normalsize{Sample Galaxies and Identified Radio Sources}}
\begin{center}
\begin{tabular}{lcccccc} \hline\hline\\[-6pt]
Galaxy    & D     & $S_{6cm}^{compact}$/$S_{6cm}^{total;a}$ & $S_{2cm}^{compact}$/$S_{2cm}^{total;a}$ & \HII$^{b}$ & SNR$^{b}$ & Indeterminated$^{b}$ \\
          & (Mpc) & (mJy/mJy) & (mJy/mJy) & Thick/Thin &     &              \\ \hline
IC 342    & 3.3   &     7.5/82          &     7.5/38          & 5/1     & 6   & 0      \\
Maffei II & 5.0   &    13.5/107         &    12.7/46          & 4/3     & 3   & 0      \\
NGC 2903  & 8.9   &     1.7/35          &     3.0/12          & 4/2     & 1   & 0      \\
NGC 6946  & 5.9   &     6.8/39          &     6.1/23          & 1/3     & 2   & 3      \\   \hline \\[-6pt]
\multicolumn{7}{l}{$^{a}$The flux ratio between the summation of compact source fluxes and total flux. Total flux is the flux} \\
\multicolumn{7}{l}{\ \ obtained from \textit{VLA} B- and C-configuration measurements (Turner \& Ho 1983; Turner \& Ho 1994; } \\
\multicolumn{7}{l}{\ \ Wynn-Williams \& Becklin 1985).} \\
\multicolumn{7}{l}{$^{b}$Number of compact sources in the category. See the definition of each category in text.}
\end{tabular}
\end{center}
\end{table*}

\section{Observations}

The radio continuum data at 2~cm and 6~cm were acquired at the \textit{NRAO} \textit{Very Large Array}{\footnote{The National Radio Astronomy Observatory is a facility of the \textit{National Science Foundation} operated under cooperative agreement by Associated Universities, Inc.}}. In order to enhance the sensitivity of these images, we have accumulate our unpublished \textit{VLA} A-configuration data with all available \textit{VLA} archival data. Only archived data sets of A-, B-, or C-configurations with time on source $>$ 10 minutes and phase center within 18 arcsec (1/10 of \textit{VLA} primary beam at 2~cm) from centers of our measurements were used. Data calibration was done by using \textit{AIPS}, following the standard reduction procedures.

The flux measurements were done with matching (\textit{u,v}) coverages at 6 and 2~cm, which include matched shortest baselines of A-configuration at 2~cm and longest baselines of A-configuration at 6~cm. The largest angular scales sampled by the images are $\sim$ $6\arcsec$. Fluxes and peak fluxes are therefore lower limits to the total flux if extended emission is present. The images were then convolved to the same beamsize, to have matching maximum baseline lengths. Final rms noise levels in blank regions of maps are $\lsim 0.05$~mJy/beam at 6~cm, and $\lsim 0.15$~mJy/beam at 2~cm. We note that the uncertainty of absolute flux scale is $\lsim 5~\%$.

\begin{figure*}[t]
\centering
%\psbox[xsize=0.4#1,ysize=0.2#1,rotate=r]
\psbox[xsize=10cm]{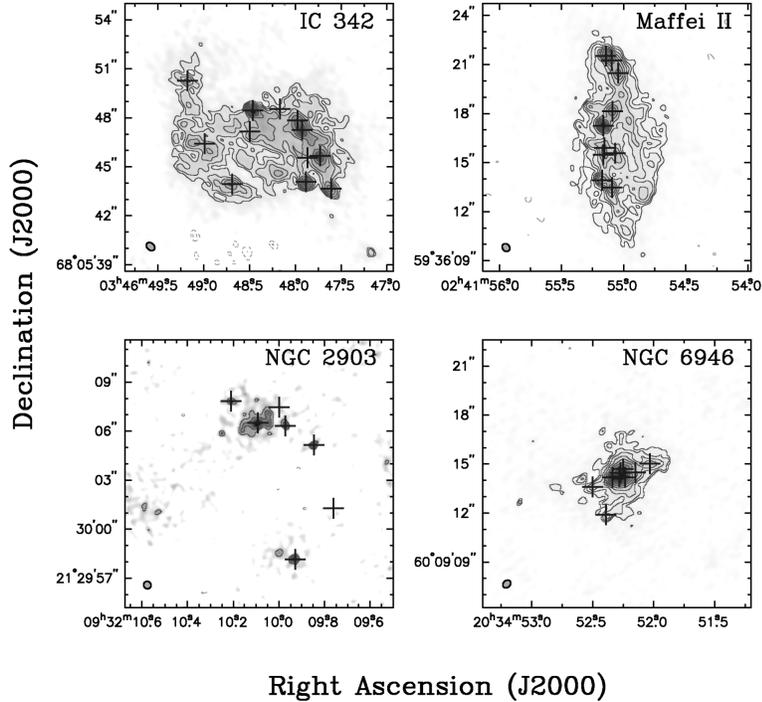}
\caption{\normalsize{\textit{VLA} 6~cm maps of our four sample galaxies: IC 342, Maffei II, NGC 2903, and NGC 6946. The maps are contoured at half-integral powers of $2 \times 0.14$~mJy/beam ($\sim 4 \sigma$) with $16\arcsec \times 16\arcsec$ field-of-view. The cross marks indicate the location of the compact radio sources which we identified at 6~cm and/or at 2~cm maps.}}
%\caption{\textit{VLA} 6~cm maps of our four sample galaxies: IC 342, Maffei II, NGC 2903, and NGC 6946. The maps are contoured at half-integral powers of $2 \times 0.14$~mJy/beam ($\sim 4 \sigma$) with $16\arcsec \times 16\arcsec$ field-of-view. The cross marks indicate the location of the compact radio sources which we identified at 6~cm and/or at 2~cm maps.}
\end{figure*}

\begin{figure*}[t]
\centering
%\psbox[xsize=0.4#1,ysize=0.2#1,rotate=r]
\psbox[xsize=10cm]{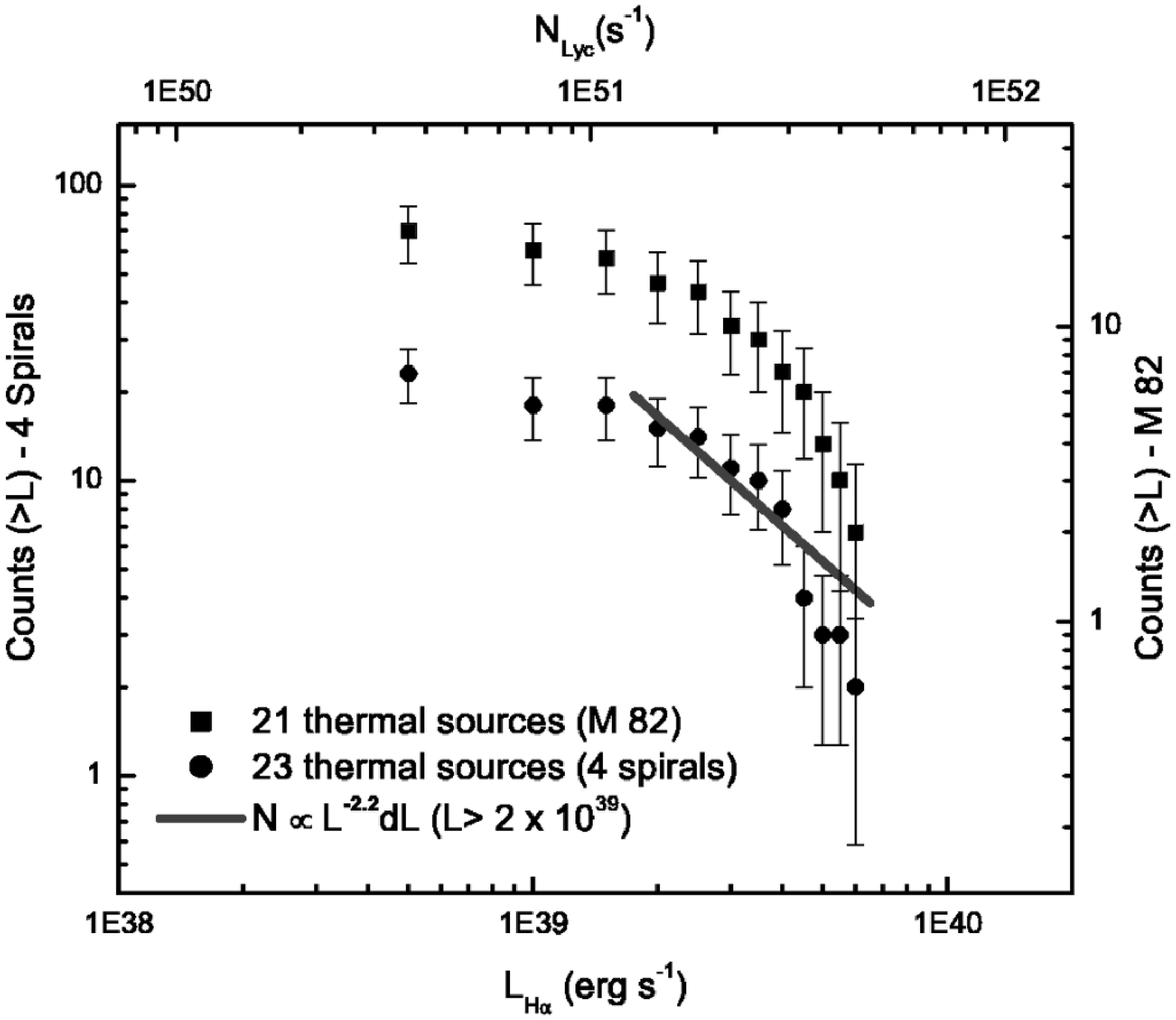}
\caption{\normalsize{Luminosity function of the radio \HII\ regions in M 82 and the centers of 4 nearby spiral galaxies: IC 342, Maffei II, NGC 2903, and NGC 6946. Filled squares and filled circles with statistical error bars show the cumulated number counts of radio \HII\ regions found in M 82 and our 4 sample spirals, respectively. The solid line represents the best-fit power-law luminosity function model for sources in 4 spirals with $L_{H_{\alpha}} > 2 \times 10^{39}~erg~sec^{-1}$. The sample is complete down to $N_{Lyc} \sim 6 \times 10^{50}~s^{-1}$ ($L_{H_{\alpha}} \sim 8.3 \times 10^{38}~erg~s^{-1}$).}}
%\caption{Luminosity function of the radio \HII\ regions in M 82 and the centers of 4 nearby spiral galaxies: IC 342, Maffei II, NGC 2903, and NGC 6946. Filled squares and filled circles with statistical error bars show the cumulated number counts of radio \HII\ regions found in M 82 and our 4 sample spirals, respectively. The solid line represent the best-fit power-law luminosity function model for sources in 4 spirals with $L_{H_{\alpha}} > 2 \times 10^{39}~erg~sec^{-1}$. The sample is complete down to $N_{Lyc} \sim 6 \times 10^{50}~s^{-1}$ ($L_{H_{\alpha}} \sim 8.3 \times 10^{38}~erg~s^{-1}$).}
\end{figure*}

\section{The Radio Continuum Images and the Compact Radio Sources}

The 6~cm naturally-weighted radio continuum maps of the four galaxies are shown in Figure 1, contoured at the same 4~$\sigma$ levels. We see emission from the central $\sim$ 150 pc except in the case of Maffei II, in which the emission extends over a distance of $\sim$ 350~pc. In each of the galaxies, the 6~cm continuum is strongest at the nucleus. Peak flux densities at 6~cm are 1.2, 2.6, 0.4, and 1.9~mJy/beam for IC 342, Maffei II, NGC 2903, and NGC 6946, respectively.

In each galaxy there are strong compact sources embedded in extended emission. The western part of the extended emission in IC 342 contains 5 sources aligned from north-west to south-east while the eastern part has fewer sources and their emission is also significantly weaker. Ten sources are identified in Maffei II, lying on a twisted, inverse ``S'' shaped line north-south over 10$\arcsec$ ($\sim$ 240~pc) on the eastern edge of the extended emission. In NGC 2903 and NGC 6949, 6 cm continuum emission is confined to single central sources surrounded by extended emission. The diffuse emission is ``patchy'' because the large scale extended emission has been resolved out by the high resolution.

Our deep, high resolution radio continuum maps at 6~cm and 2~cm reveal 38 compact radio sources (cross marks in Figure 1). The sources meet at least one of the following criteria: (1) $5~\sigma$ detection at the peak, (2) $5~\sigma$ in integrated emission at one wavelength, or (3) $4~\sigma$ emission detection in both wavebands. 

The spectral index, defined as $S_{\nu} \propto \nu^{\alpha}$, indicates the nature of a radio source. Based on the values of $\alpha_{6-2}$, we classified sources as \HII--thick (optically thick free-free -- $\alpha > 0.0$), \HII--thin (optically thin free-free -- $\alpha \sim -0.1$), SNR (supernova remnant; synchrotron -- $\alpha < -0.1$). For the cases identified as \HII\ regions, the required Lyman continuum rate, $N_{Lyc}$, can be derived from the observed flux density at 2 cm. $N_{O7}$, the number of standard O7 stars needed to generate $N_{Lyc}$ in a radio nebula, can be also derived. We here assume ionization-bonded nebulae; if UV photons escape the \HII\ region, our OB luminosities will be underestimates of the true Lyman continuum rates.

\section{Discussion}

More than half of the 38 identified compact sources are \HII\ regions with flat or positive indices ($\alpha \gsim -0.1$). More than 60\,\% of these \HII\ regions have a spectrum rising from 6 to 2 cm, indicating that they are at least partly optically thick at 6~cm. The radio fluxes of these radio \HII\ regions require that they harbor hundreds of massive stars. For electron temperatures of $\sim 10,000$~K, rising cm-wave spectra imply that the rms electron densities are $\sim 10^{4}~cm^{-3}$.

The deconvolved sizes of \HII\ regions correspond to diameters of $\sim$ 6 - 14~pc at the distances of the galaxies. These \HII\ regions are much smaller than the 30 Doradus \HII\ region ($\sim$ 200~pc in diameter for region of $EM > 10^{4}~pc~cm^{-6}$; see Mills et al. 1978 and Kennicutt 1984) in the Large Magellanic Cloud (LMC). They are larger but similar in nature to the Galactic compact \HII\ regions in W49 (Mezger et al. 1967; Conti \& Blum 2002) and NGC 3603 (de Pree et al. 1999; Sung \& Bessell 2004, M{\" u}cke et al. 2002). One possibility for the fact that these \HII\ regions are so compact and dense is that they are younger than 30 Doradus or NGC 3603.

Nearly two dozen radio sources in the four galaxies of this study were found to have flat or rising spectra. These are presumably thermal nebulae excited by massive stars, for which the 2 cm fluxes give a lower limit to the required $N_{lyc}$ and corresponding $L_{H_{\alpha}}$. We cannot detect unresolved (diameter $<$ 15 pc) nebulae with $N_{Lyc} < 4\times 10^{50}~ s^{-1}$ due to our sensitivity limits, corresponding to 40 O stars in IC 342 and 120 stars in NGC 6946. However, we also cannot detect more luminous regions if they are large and resolved (size $\gg$ 15~pc); in this limit we are sensitive to \HII\ regions with $EM > 10^4 \rm ~cm^{-6}\,pc$. The youngest and therefore smallest \HII\ regions are probably not affected by this latter criterion.

The luminosity function (LF) of the 23 thermal sources in 4 spirals shown in Figure 2 suggests a broken power law with turnover point at $L_{H_{\alpha}} = 2 \times 10^{39} erg/sec$ and cutoff at luminous end around $L_{H_{\alpha}} = 10^{40} erg/sec$. The total mass of such a cluster at cutoff would be $\sim 2 \times 10^{5} \; M_{\sun}$ (assuming Salpeter initial mass function, IMF), consistent with the mass of Galactic globular clusters. Very similar feature shows in LF for 21 thermal radio sources in M 82, another nearby starburst galaxy, based on more complete radio studies of Allen (1999), with sensitivity limits of $N_{Lyc} < 2\times 10^{50}~ s^{-1}$ (20 O7 stars) and radio nebulae with size $\lsim$ 50 pc. The agreement of the LF shape of M 82 and our sample suggests that the broken power law may be real. However, the results based on luminosity function of 44 sources are statistically inconclusive. Larger samples would help to clarify this issue.

The steep power law LF suggests a cutoff at the high luminosity end in the young clusters associated with radio nebulae in spiral galaxies. The upper end of the luminosity function is around $L_{H_{\alpha}} \sim 10^{40}~ erg~ s^{-1}$. This corresponds to $\sim$ 750 O7 stars. We observe two such clusters, in Maffei 2 and NGC 2903, and their nebulae are $\lsim5$ pc in extent. The total mass of such a cluster would be $\sim$ 2.2 $\times 10^{5}~ M_{\sun}$ (assuming Salpeter IMF), consistent with the mass of Galactic globular clusters.

\section{Summary}

We have identified 38 compact radio emission in the centers of four nearby spiral galaxies, IC 342, Maffei II, NGC 2903, and NGC 6946, at subarcsecond resolution at 6 and 2 cm with the VLA. Over half of the compact sources appear to be \HII\ regions, based on their radio spectra, of which two-thirds appear have optically thick free-free emission. The compact \HII\ regions contain ionized gas with densities as high as $n_{i} \sim 10^{4}~ cm^{-3}$, which suggests that these \HII\ regions are relatively young. The largest \HII\ regions we detect require equivalent of $\sim$ 500 - 1,000 O7 stars to excite them; since they are dense, compact \HII\ regions, by analogy with Galactic compact \HII\ regions, they are likely to be very youngest of the massive young clusters, a Myr or less in age.

\section*{acknowledgments}

We acknowledge support by U.S. NSF grants AST 0307950 and AST 0071276 for this research.

\section*{References}

\re Allen, M.~L. 1999, Ph.D.~Thesis, Univ. Toronto
\re Beck, S.~C. et al. 1996, ApJ, 457, 610
\re B{\" o}ker, T. et al. 2002, AJ, 123, 1389
\re Conti, P.~S. \& Blum, R.~D. 2002, ApJ, 564, 827
\re de Pree, C.~G. et al. 1999, AJ, 117, 2902
\re Ho, P.~T.~P. et al. 1990, ApJ, 349, 57
\re Holtzman, J.~A. et al. 1992, AJ, 103, 691
\re Kawara, K. et al. 1989, ApJ, 337, 230
\re Kennicutt, R.~C. 1984, ApJ, 287, 116
\re Lada, C.~J. \& Lada, E.~A. 2003 ARA\&A, 41, 57
\re Larson, R.~B. 1992, in ``The Astronomy and Astrophysics Encyclopedia'', ISBN 0-442-26364-3, page 672
\re Mezger, P.~G. et al. 1967, ApJ, 150, 807
\re Mills, B.~Y. et al. 1978, MNRAS, 185, 263
\re M{\" u}cke, A. et al. 2002, ApJ, 571, 366 
\re Sung, H. \& Bessell, M.~S. 2004, AJ, 127, 1014
\re Turner, J.~L. \& Ho, P.~T.~P. 1983, ApJl, 268, L79
\re Turner, J.~L. \& Ho, P.~T.~P. 1994, ApJ, 421, 122
\re Tsai, C.-W. et al. 2006 AJ, submitted
\re Whitmore, B.~C. 2003, STScI symp. series, 14, 153
\re Wynn-Williams, C.~G. et al. 1972, MNRAS, 160, 1
\re Wynn-Williams, C.~G. \& Becklin, E.~E. 1985, ApJ, 290, 108
\end{document}